\documentclass[11pt]{article}

\usepackage[preprint]{acl}
\usepackage{times}
\usepackage{latexsym}

\usepackage[T1]{fontenc}
\usepackage[utf8]{inputenc}

\usepackage{microtype}
\usepackage{inconsolata}

\usepackage{graphicx}
\usepackage{booktabs}
\usepackage{multirow}
\usepackage{tabularx}
\usepackage{amsmath}
\usepackage{amssymb}
\usepackage{mathtools}
\usepackage{subcaption}
\usepackage{enumitem}
\usepackage{xcolor}
\usepackage{url}
\usepackage{cleveref}

\newcommand{\ie}{\textit{i.e. }}
\newcommand{\eg}{\textit{e.g. }}

\title{MTRACE: Multilingual Retrieval-Augmented Generation for Temporally Diverse Text Corpora}

\author{Souhail Bakkali$^{1,\dagger}$, Anthony Mudet$^{2}$, \\
$^{1}$Univ Rennes, CNRS, IRISA - UMR 6074, Rennes, France\\
$^{2}$L3i-lab, La Rochelle Université, France\\
{\tt\small souhail.bakkali@irisa.fr},
{\tt\small anthonymudet94@gmail.com} 
}

\begin{document}
\maketitle
\begin{abstract}
Large multilingual knowledge bases expose temporally diverse information, yet retrieval quality remains sensitive to lexical variation and cross-lingual terminology shifts. We develop and evaluate MTRACE (Multilingual Temporal Retrieval-Augmented Generation with evidence grounding), a pipeline designed to test whether query expansion and multi-query fusion mitigate vocabulary mismatch in temporally layered text corpora, on the French and English subsets of MIRACL. Our approach integrates: (i) semantic query expansion (SQE) and multi-query fusion via Reciprocal Rank Fusion (RRF), targeting retrieval stability under query variation; (ii) a generation prompt enforcing strict grounding in retrieved evidence and explicit abstention when evidence is insufficient; and (iii) a modular architecture enabling systematic component evaluation. Ablation studies on Named Entity Recognition (NER) and embedding model selection demonstrate the importance of syntactic coherence in entity extraction and of self-retrieval and efficiency measurements for retriever selection. Our end-to-end evaluation over 50 constructed queries shows faithful answers for well-supported queries, correct abstention on unanswerable questions, and no re-scored similarity gains from multi-query fusion over single-query dense retrieval. By scoping our claims to a clean, text-only baseline, we separate these effects from OCR-noise confounds; direct measurement of diachronic lexical drift is left to future work. Code and configurations are available at \url{https://anonymous.4open.science/r/MIRAGE-8EAA/}.
\end{abstract}

\section{Introduction}
\label{sec:intro}

Large multilingual knowledge bases such as Wikipedia carry immense temporal depth, valuable for longitudinal analysis of socio-cultural trends, linguistic evolution, and historical events. Yet their computational exploitation for open-domain question answering is hampered by diachronic language change and cross-lingual heterogeneity: unlike static, contemporary corpora, temporally layered text exhibits semantic drift, where named entities, geopolitical designations (\eg, ``Burma'' to ``Myanmar''), and common phrases evolve over decades of editorial activity.

General-purpose multilingual text corpora disrupt modern NLP pipelines through: (i) lexical and orthographic variation from diachronic change, which reduces overlap between contemporary queries and archived passages; (ii) named entity evolution, where persons and organizations acquire different designations over time \citep{pan-etal-2017-cross,tedeschi2021wikineural}; (iii) structural heterogeneity from evolving editorial and template conventions; and (iv) domain shift between encoder pretraining corpora and temporally stratified encyclopedic content. Consequently, sparse and dense retrieval methods that assume stable vocabulary distributions frequently suffer vocabulary mismatch on queries requiring temporal reasoning \citep{schweter2022hmbert,reimers-2020-multilingual-sentence-bert}.

Retrieval work has addressed language-specific encoders and dense passage architectures \citep{zhang2023miracl}. Retrieval-Augmented Generation (RAG) \citep{lewis2020retrieval} grounds LLM responses in retrieved evidence, but its quality critically depends on retrieval, which remains brittle under the lexical and semantic variation endemic to temporally diverse text. Prior RAG research has focused on static, web-scale corpora or heavily degraded historical archives, so relatively little work evaluates RAG where OCR noise can be held constant. We use the MIRACL dataset \citep{zhang2023miracl} as a controlled testbed: on a comparatively clean, multilingual corpus we remove severe OCR noise and layout artifacts as confounds, and study how retrieval behaves under lexical and cross-lingual variation in temporally layered content.

Our contributions are:
\begin{itemize}
    \item MTRACE, a systematic, text-only RAG pipeline for multilingual QA over temporally layered text, designed to mitigate vocabulary mismatch under lexical and cross-lingual variation.
    \item Empirical validation of component choices through ablation studies, showing the importance of syntactic coherence in NER and of self-retrieval and efficiency diagnostics for embedding model selection.
    \item Quantitative characterization of hybrid retrieval on French and English MIRACL: under similarity-based re-scoring across four embedding models, multi-query fusion does not outperform single-query dense retrieval (we analyze why), while structured generation produces faithful answers with appropriate abstention; scope and limitations are stated explicitly.
\end{itemize}

\section{Related Work}
\label{sec:related}

\subsection{Retrieval-Augmented Generation}
RAG \citep{lewis2020retrieval} grounds LLMs in external knowledge sources, reducing factual hallucination and improving the verifiability of generated text: a retriever fetches documents that are used as context by a generator. Extensions include iterative retrieval \citep{trivedi2023interleavingretrievalchainofthoughtreasoning}, fused token-level generation \citep{izacard2021leveragingpassageretrievalgenerative}, and retriever fine-tuning \citep{izacard2022atlasfewshotlearningretrieval}. A critical limitation is the dependence on the initial retrieval step: when the retriever misses relevant evidence, the generator has little chance of producing a correct answer \citep{vidgen2020directions}. This sensitivity is exacerbated by lexical variation, linguistic drift, and multilingual complexity---precisely the conditions present in MIRACL \citep{zhang2023miracl}. While RAG has succeeded on clean, static corpora, its application to multilingual and temporally diverse collections remains relatively unexplored, particularly when queries and passages span different time periods or languages. On the retrieval side, multi-query fusion via RRF \citep{rackauckas2024ragfusion} and its deployment-scale follow-up \citep{medrano2026ragfusion} exemplify robustness patterns that recent surveys and empirical best-practice studies catalog alongside re-ranking and self-reflective retrieval \citep{gao2023retrievalaugmented,wang2024searching,asai2024selfrag}.

\subsection{NLP for Temporally Diverse and Multilingual Corpora}
Applying NLP to corpora with significant temporal depth is well documented. Digital humanities projects such as NewsEye \citep{doucet2020newseye} and impresso \citep{ehrmann2020language} pioneered NER \citep{10.1007/978-3-030-58219-7_21}, topic modeling \citep{hengchen2021challengescomputationallexicalsemantic,Blevins2015JaneJ}, and article segmentation \citep{10.1145/3340531.3412767} for historical newspaper collections; the core challenges they identified, entity evolution and orthographic variation, are equally present in the temporal depth of encyclopedic corpora like Wikipedia, whose editorial evolution MIRACL inherits. Parallel research addresses diachronic language directly, from orthographic normalization \citep{dong2018multi} and historical language models like HMBERT \citep{schweter2022hmbert} to lexical-semantic-change measurement \citep{schlechtweg2020semeval,tahmasebi2018survey}. Recent benchmarks such as MIRACL \citep{zhang2023miracl} and the CLEF HIPE shared tasks \citep{10.1007/978-3-030-58219-7_21,ehrmann2022overview} establish multilingual evaluation frameworks, yet a significant gap remains in integrating robust retrieval and generation into cohesive pipelines for open-domain QA over multilingual corpora with a significant temporal dimension.

\subsection{Robust Information Retrieval}
A core challenge is performing effective retrieval despite lexical and semantic variation across time and languages. Query expansion---leveraging word embeddings to identify conceptually related terms beyond lexical overlap \citep{roy2016usingwordembeddingsautomatic,Carpineto2012ASO}---augments the query to mitigate vocabulary mismatch, which is especially relevant when contemporary queries differ from the language of older articles. Hybrid strategies \citep{thakur2021beirheterogenousbenchmarkzeroshot} combine sparse (\eg, BM25) and dense retrievers; learned sparse expansion \citep{formal2021splade}, late interaction \citep{khattab2020colbert}, and hypothetical-document expansion \citep{gao2023hyde} offer further routes; we restrict our study to multi-query fusion for tractability. RRF \citep{10.1145/1571941.1572114} aggregates ranked lists without score calibration, making it well suited to combining heterogeneous retrievers across languages. We investigate query expansion with RRF as a resilient front end for multilingual QA that bridges terminology gaps from cross-lingual variation and temporal drift. \citet{gonzalez2024retrieval} pioneered RAG for historical texts with a dense retriever; we adopt their core insight---that RAG can bridge user queries and archival content---while extending it to the multilingual, temporally diverse setting of MIRACL through hybrid retrieval.

\section{Methodology}
\label{sec:rag}

\begin{figure*}[t]
    \centering
    \includegraphics[width=\linewidth]{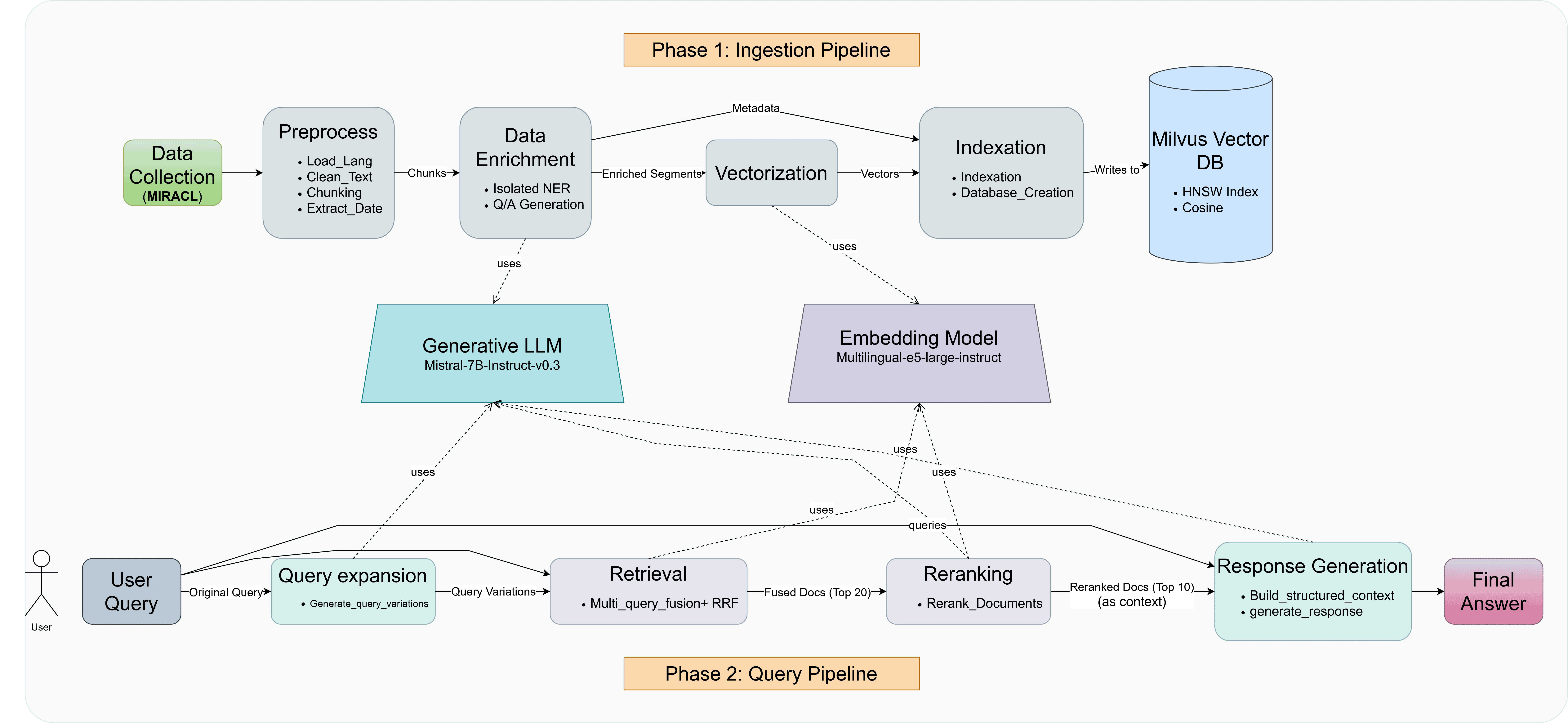}
    \caption{Overview of MTRACE, our multilingual RAG pipeline. \textbf{Ingestion (Phase 1):} preprocessing, chunking, NER, and vector indexing. \textbf{Query (Phase 2):} the query is expanded via an LLM, the original and reformulated query rankings are fused with RRF, and a generator constrained to retrieved evidence answers the query, abstaining when evidence is insufficient.}
    \label{fig:pipeline}
\end{figure*}

We build a three-stage pipeline, \textbf{MTRACE} (\textbf{M}ultilingual \textbf{T}emporal \textbf{R}etrieval-\textbf{A}ugmented \textbf{G}eneration with \textbf{E}vidence grounding; \cref{fig:pipeline}): retrieval over the preprocessed corpus (\cref{subsec:retrieval}), structured generation from the retrieved evidence (\cref{subsec:generation}), and a modular design that lets us ablate each stage separately (\cref{sec:results}).

\subsection{Problem Formulation}
\label{subsec:problem_formulation}

We define text QA over a corpus \(\mathcal{C} = \{d_1, \ldots, d_N\}\) of chunks exhibiting diachronic lexical variation, entity-name evolution, and cross-lingual terminology shifts. Given a query \(q\) expressed in natural language, we \textbf{Retrieve} an ordered set \(R = \{d_{r_1}, \ldots, d_{r_k}\} \subset \mathcal{C}\) with high relevance to \(q\), and \textbf{Generate} a fluent, faithful answer \(a\) that addresses \(q\) while being grounded exclusively in the evidence provided by \(R\).
We optimize \(a^* = \arg\max_a P(a \mid q, R)\), subject to \(\text{Faithfulness}(a, R) \geq \tau\), where \(\tau\) is an informal faithfulness requirement---we report faithfulness as a continuous RAGAS score (\cref{subsec:metrics}) rather than enforcing a hard constraint. The retrieved set is \(R = \text{Retrieve}(q, \mathcal{C}; \theta_R)\), with \(\theta_R\) the retriever's parameters; grounding must hold despite the temporal heterogeneity and cross-lingual disparities of chunks in \(R\).

\subsection{Hybrid Retrieval Module}
\label{subsec:retrieval}

\subsubsection{Dense Retriever and Query Expansion}
Our dense retriever is \texttt{multilingual-e5-large-instruct}, selected through the ablation in \cref{subsec:exp_embedding} for its balance of cross-lingual semantic understanding, retrieval quality, and computational efficiency among five models; its instruction tuning suits the complex semantic relationships of temporally layered text.

To address vocabulary mismatch between contemporary queries and historical or cross-lingual archived wording, we expand each query \(q\) with the \texttt{mistralai/Mistral-7B-Instruct-v0.3} LLM, prompted (\cref{tab:prompt_query_variation}) for \(N=5\) semantically equivalent but lexically diverse reformulations \(Q' = \{q_1, \ldots, q_5\}\). The reformulations are not constrained by content category; the rank-fusion step in \cref{subsec:retrieval} aggregates whichever lexical perspectives the model produces.

We treat this expansion-and-fusion pipeline as the hypothesis under test rather than an assumed improvement: \cref{subsec:exp_e2e} evaluates it against single-query dense retrieval under a controlled re-scoring protocol. The full set \(Q = \{q\} \cup Q'\) searches the dense index of the preprocessed corpus \(\mathcal{C}'\) in parallel, casting a wider net over lexical variants introduced by historical language evolution---compensating for the retriever's sensitivity to exact term matching when temporal or cross-lingual variation shifts subword tokenization in document representations.

\subsubsection{Reciprocal Rank Fusion for Robust Ranking Aggregation}
Ranked lists from each query in the expanded set \(Q\) are aggregated with RRF \citep{10.1145/1571941.1572114}, which operates solely on document ranks---robust to the incompatible similarity scales of neural retrievers across query variants, unlike score-based fusion that requires calibration.

The RRF score for a document \(d\) is \(\text{RRF}(d) = \sum_{q_i \in Q} \frac{1}{k + \text{rank}(d, q_i)}\), where \(\text{rank}(d, q_i)\) denotes the position of document \(d\) in the ranked results for query variation \(q_i \in Q\), and \(k\) is a smoothing hyperparameter (empirically \(k = 60\)) controlling the influence of lower-ranked documents. RRF mitigates poorly-formulated expansions by aggregating multiple perspectives, combines heterogeneous retrievers without score normalization (rank invariance), boosts documents appearing consistently across lists, and matches different surface forms of the same semantic content.

The final retrieved set \(R\) is constructed by selecting the top-\(K\) documents after re-ranking by their aggregated RRF scores (we use \(K=10\)). We evaluate the effect of this multi-query fusion against single-query retrieval in \cref{subsec:exp_e2e}.

\subsection{Augmented Generation Module}
\label{subsec:generation}

\subsubsection{Context Structuring and Evidence Organization}
We structure retrieved chunks \(R\) with three operations that enhance reasoning and attribution: (1) \textbf{source grouping} merges chunks from identical articles to prevent fragmentation; (2) \textbf{metadata enrichment} prepends article titles and document identifiers, providing temporal and provenance context; and (3) \textbf{visual delineation} separates sources with clear markers ("\texttt{\textbackslash n\textbackslash n---\textbackslash n\textbackslash n}") to distinguish potentially contradictory information, supporting claim traceability and chronological disambiguation.

\subsubsection{Constrained Generation via Prompt Engineering}
We employ \texttt{Mistral-7B-Instruct-v0.3} (temperature 0.3) with a prompt enforcing strict evidence grounding through four constraints: retrieved context as the sole permitted source; abstention when evidence is insufficient; same-language responses (\cref{tab:prompt_historical_qa}); and explicit verification of entity relationships against the provided context. Full templates are given in \cref{tab:prompt_qa,tab:prompt_historical_qa}.

\section{Experiments and Results}
\label{sec:results}

\subsection{Dataset and Preprocessing}
\label{subsec:data_preprocessing}

We use a subset of MIRACL (Multilingual Information Retrieval Across a Continuum of Languages) \citep{zhang2023miracl}, which comprises Wikipedia passages in 18 languages with human-annotated relevance judgments. The clean, text-only baseline removes low-quality digitization and OCR errors as confounds, leaving lexical and cross-lingual variation in temporally layered content as the factors under study.

\subsubsection{Corpus Composition}
Our subset focuses on French and English passages, totaling 688,992 text chunks. Each document includes a title and body text; chunks are obtained by sentence-boundary segmentation of each article, followed by removal of exact duplicate sentences. The multilingual nature of the corpus enables evaluation of cross-lingual retrieval capabilities.

\subsubsection{Preprocessing Pipeline}
We apply a minimal cleaning procedure---Unicode normalization, HTML tag removal, and whitespace/punctuation correction---to ensure a consistent baseline. The resulting standardized corpus $\mathcal{C'}$ maintains the original content while removing trivial formatting inconsistencies.

\subsubsection{Splits and Evaluation Queries}
We adopt the standard MIRACL train/test splits. For qualitative analysis and RAGAS assessment, we construct 50 evaluation queries spanning three categories: (i) entity-focused questions (\eg, {\ttfamily "Who was Antoine Meillet?"}), (ii) event-oriented queries (\eg, {\ttfamily "What caused the American Civil War?"}), and (iii) intentionally unanswerable questions to evaluate abstention behavior---20/20/10, evenly split between French and English (25 each). The queries were written by the authors against the MIRACL French and English content without reference to any system output (no query was selected on the basis of retrieval or generation results). The complete set, with per-category counts, is released at \url{https://anonymous.4open.science/r/MIRAGE-8EAA/}; per-query results in \cref{subsec:exp_e2e} are aggregated over these 50 queries.

\subsubsection{Implementation Details:} The corpus is indexed as 688{,}992 chunks of 512 tokens with the dense retriever only (no sparse index is deployed); retrieval feeds the generator \(K=10\) candidates fused by RRF (\(k=60\)) at temperature 0.3, and each query entails two generator calls: one producing the five reformulations, one producing the answer. Timing provenance differs by experiment---encoding times in \cref{tab:embedding-comprehensive} were measured in the original cluster environment, while re-ranking times in \cref{tab:benchmark_retrieval_fusion_detail} and answer generation behind \cref{tab:ragas_results_detailed} used a single NVIDIA T4 (16\,GB) in fp16. \cref{sec:app_impl} collects the full hardware, precision and configuration record, including the 4-bit re-scoring used for the appendix baselines.

\subsection{Evaluation Metrics}
\label{subsec:metrics}

We score retrieval with \textbf{Sim@1} and \textbf{Sim@5}, the mean cosine similarity between a query embedding and its top-1/top-5 re-scored candidates, and the \textbf{confidence drop} (the top-1/top-2 gap); embedding selection additionally uses the self-retrieval variant of Sim@5, in which each chunk queries the index including itself. NER is scored with \textbf{Syntactic Relevance}, the fraction of extracted entities that form complete linguistic units. Generation is scored with the RAGAS framework \citep{es2024ragas}: \textbf{faithfulness}, the fraction of an answer's claims supported by the retrieved context, and \textbf{answer relevancy}, the cosine agreement between the answer and the query via synthetic questions. Formal definitions, the self-retrieval constant, and the clustering diagnostics (Silhouette, Davies--Bouldin, Calinski--Harabasz) are given in \cref{sec:app_metrics}; in the reported runs, claim verification uses \texttt{deepseek-v4-flash} and relevancy embeddings use \texttt{qwen3-embedding:4b}, hosted stand-ins for the RAGAS defaults.

\subsection{Ablation Component: Entity-Aware Retrieval Enrichment (NER)}
\label{subsec:exp_ner}

\subsubsection{Experimental Setup:} We evaluated four NER models on a balanced sub-corpus of 50,000 English and French MIRACL texts, judging three dimensions for downstream retrieval augmentation: \textit{processing efficiency}, \textit{entity extraction volume}, and \textit{syntactic relevance}---the ability to extract entities that form coherent linguistic units without fragmentation or boundary errors. Syntactic coherence is a prerequisite for effective query expansion: fragmented entities introduce noise that degrades downstream retrieval.

\subsubsection{Comparative Analysis:} \cref{fig:ner-comparison} shows clear trade-offs. \texttt{bert-base-multilingual-cased} yields the highest raw entity count (\cref{fig:ner-entities}) but produces frequent fragmentation (\eg, `\#\#iste allemand Walter Porzig`) and inconsistent labels (`LABEL\_0`, `LABEL\_1`; \cref{tab:ner-qualitative-comparison}), noise that would degrade downstream entity-aware retrieval. The historically-trained \texttt{bert-base-historic-multilingual-cased} underperformed---neither improved accuracy nor better handling of archaic spellings, while incurring computational overhead---suggesting that domain-specific pretraining is insufficient without explicit optimization for the NER task.

\begin{table}[t]
    \centering
    \caption{Qualitative comparison of entity extraction.}
    \label{tab:ner-qualitative-comparison}
    \resizebox{0.9\columnwidth}{!}{%
    \begin{tabular}{|@{}p{0.3\linewidth}|p{0.66\linewidth}@{}|}
        \hline
        \textbf{Model} & \textbf{Extracted Entity Examples} \\
        \hline
        \texttt{wikineural} & 
        \texttt{[('Walter Porzig', 'PER'), ('Mei', 'PER')]} \\
        \texttt{bert-base-\allowbreak multilingual-\allowbreak cased} & 
        \parbox[t]{\dimexpr0.74\linewidth-2\tabcolsep\relax}{\ttfamily 
        [('Selon le lingu', 'LABEL\_1'),\\
        ('\#\#iste allemand Walter Porzig', 'LABEL\_0'), ...]
        } \\
        \hline
    \end{tabular}
    }%
\end{table}

\subsubsection{Model Selection Justification:} We select \texttt{wikineural-multilingual-ner} as the optimal balance across our criteria: high syntactic relevance (85\%, \cref{fig:ner-syntax}) at competitive speed (\cref{fig:ner-speed}), and clean, well-typed outputs (`('Walter Porzig', 'PER')`) directly usable for entity-based query expansion without post-processing.

\subsection{Ablation Component: Embedding Model Selection and Efficiency Trade-offs}
\label{subsec:exp_embedding}

\subsubsection{Experimental Setup:} We evaluate five embedding models on the full corpus of 688,992 multilingual text chunks along three dimensions, following the multi-criteria spirit of embedding benchmarks such as MTEB \citep{muennighoff2023mteb}: (i) semantic retrieval performance, measured in the self-retrieval setting described in \cref{subsec:metrics}; (ii) computational efficiency; and (iii) embedding-space diagnostics.

\begin{table}[t]
    \centering
    \caption{Comprehensive evaluation of embedding models (self-retrieval protocol; \texttt{multilingual-e5-large} denotes \texttt{intfloat/multilingual-e5-large-instruct}).}
    \label{tab:embedding-comprehensive}
    \resizebox{0.9\columnwidth}{!}{%
        \begin{tabular}{|l|l|l|l|l|l|l|}
        \hline
            \textbf{Model} & \textbf{Top-5} & \textbf{Drop} & \textbf{Time(s)} & \textbf{Dim.} & \textbf{Semantic} & \textbf{Efficiency} \\
        \hline
            \texttt{e5-small-v2} & 0.9073 & 0.105 & 125 & 384 & Excellent & Good compromise \\
            \texttt{multilingual-e5-large} & 0.9134 & 0.097 & 360.6 & 1024 & Excellent & Average \\
            \texttt{SFR-Embedding-Mistral} & 0.8123 & 0.2090 & 6143 & 4096 & Good & Long + heavy \\
            \texttt{linq-embed-mistral} & 0.7410 & 0.2918 & 6184 & 4096 & Fairly good & Long + heavy \\
            \texttt{MiniLM} & 0.7222 & 0.3087 & 72.91 & 384 & Low & Very fast/lightweight \\
        \hline
        \end{tabular}
        }%
\end{table}

\subsubsection{Semantic Performance Analysis:} \cref{tab:embedding-comprehensive} shows a clear hierarchy: the E5 family (\ie \texttt{e5-small-v2} and \texttt{multilingual-e5-large}) reaches Top-5 similarity above 0.90 versus 0.72--0.81 for the remaining models, with smaller top-1/top-2 drop (0.097--0.105)---their nearest non-self neighbours are closer to the query chunk, the geometry signal we use for selection alongside efficiency.

\subsubsection{Efficiency and Scalability Considerations:} Efficiency gaps are dramatic: \texttt{SFR-Embedding-Mistral} and \texttt{linq-embed-mistral} need 6140\,s to encode the corpus---50$\times$ slower than \texttt{e5-small-v2} and 17$\times$ slower than \texttt{multilingual-e5-large}---impractical for iterative, large-scale deployment (\cref{tab:embedding-comprehensive}).

\subsubsection{Selection Rationale and Clustering Diagnostics:} We select \texttt{multilingual-e5-large} for its highest self-retrieval Top-5 similarity (0.9134) at acceptable encoding cost (\cref{tab:embedding-comprehensive}); the instruction-tuned variant, \texttt{intfloat/multilingual-e5-large-instruct}, is the dense retriever used in \cref{subsec:retrieval}. \texttt{e5-small-v2} is faster (125s vs.\ 360.6s) but scores lower on Top-5 (0.9073). The clustering diagnostics in \cref{tab:embed-cluster} fit $k$-means ($k=10$, \texttt{random\_state}=42) on the document embeddings, compute Silhouette on a 10,000-document subsample (\texttt{random\_state}=42), and derive Davies-Bouldin and Calinski-Harabasz from the assignments. All Silhouette values are near zero ($-0.001$ to $0.014$), indicating strongly overlapping clusters for every model, so we treat these diagnostics as a weak negative filter rather than evidence of well-separated semantic structure (\texttt{multilingual-e5-large} has the highest Calinski-Harabasz, 10430.90, and a competitive Davies-Bouldin of 5.66); retriever selection rests on the self-retrieval and efficiency results in \cref{tab:embedding-comprehensive}.

\begin{table}[t]
    \centering
    \caption{Clustering diagnostics of the vector space.}
    \label{tab:embed-cluster}
        \resizebox{0.9\columnwidth}{!}{%
        \begin{tabular}{|l|c|c|c|}
        \hline
        \textbf{Model} & \textbf{Silhouette} & \textbf{DB ($\downarrow$)} & \textbf{CH ($\uparrow$)} \\
        \hline
        \texttt{multilingual-e5-large} & 0.014 & 5.66 & 10430.90 \\
        \texttt{MiniLM} & 0.01 & 6.13 & 10286.34 \\
        \texttt{SFR} & 0.0106 & 6.15 & 7032.89 \\
        \texttt{e5-small-v2} & -0.001 & 5.08 & 7627.89 \\
        \texttt{linq-mistral} & 0.01 & 6.69 & 9065.89 \\
        \hline
    \end{tabular}
    }%
\end{table}

\subsection{Integrated RAG Pipeline Performance}
\label{subsec:exp_e2e}

\subsubsection{Impact of Hybrid Retrieval Strategy}
MTRACE's hybrid retrieval stage (query expansion followed by multi-query fusion) is evaluated against single-query dense retrieval. Protocol (\cref{tab:benchmark_retrieval_fusion_detail}): the dense condition retrieves top-\(K\) candidates for the original query with \texttt{multilingual-e5-large-instruct}; the fusion condition repeats retrieval for the query and its reformulations and fuses the results; each listed model then re-encodes the query and its candidate set, and we report mean cosine Sim@1/Sim@5 and the top-1/top-2 gap, averaged over the 50 evaluation queries (\cref{subsec:metrics}). Fusion lowers query-centric similarity for all four models: Sim@1 changes by \(-0.010\) for \texttt{multilingual-e5-large} (bootstrap 95\% CI \([-0.015,-0.006]\)), \(-0.087\) for \texttt{e5-mistral-7b-instruct} (\([-0.122,-0.054]\)), \(-0.022\) for \texttt{SFR-Embedding-Mistral} (\([-0.034,-0.012]\)), and \(-0.030\) for \texttt{Linq-Embed-Mistral} (\([-0.044,-0.018]\)); Sim@5 also decreases for all four (all CIs exclude zero), and the top-1/top-2 gap widens in every row. Part of the pattern follows from the metric itself: both conditions are re-scored against the \emph{original} query embedding, whereas RRF deliberately retains candidates that rank well under a reformulation, penalising candidates favoured by a reformulation but less aligned with the original surface form---and for \texttt{multilingual-e5-large}, retriever and rescoring model coincide, making its dense list the argmax by construction. The differences are small for three of the four models but consistent across queries, so we do not read fusion as improving re-scored similarity under this protocol.

\begin{table*}[t]
    \centering
    \footnotesize
    \caption{Single-query dense retrieval (D) vs.\ multi-query fusion (F): mean cosine Sim@1/Sim@5 and top-1/top-2 gap when each condition's top-\(K\) candidates are re-scored by the listed model, averaged over the 50 evaluation queries. Time (s) is total re-ranking time over the 50 queries under both conditions (model load excluded), measured on an NVIDIA T4 in fp16.}
    \label{tab:benchmark_retrieval_fusion_detail}
    \resizebox{\textwidth}{!}{%
    \begin{tabular}{|p{4cm}|lll|lll|l|}
        \hline
        Model & @1 (D) & @5 (D) & $\Delta 1 \rightarrow 2$ (D) & @1 (F) & @5 (F) & $\Delta 1 \rightarrow 2$ (F) & Time (s) \\
        \hline
        \texttt{multilingual-e5-large} & 0.8753 & 0.8645 & 0.0085 & 0.8653 & 0.8505 & 0.0102 & 5.0 \\
        \texttt{e5-mistral-7b-instruct} & 0.6796 & 0.5729 & 0.0707 & 0.5928 & 0.4685 & 0.0776 & 46.3 \\
        \texttt{SFR-Embedding-Mistral} & 0.7208 & 0.6863 & 0.0252 & 0.6987 & 0.6604 & 0.0263 & 49.2 \\
        \texttt{Linq-Embed-Mistral} & 0.6080 & 0.5679 & 0.0247 & 0.5777 & 0.5344 & 0.0265 & 50.4 \\
        \hline
    \end{tabular}
    }
\end{table*}

\begin{table*}[t!]
    \centering
    \footnotesize
    \caption{RAGAS evaluation results with generated answers (4 representative queries of 50; categories as in \cref{subsec:data_preprocessing}).}
    \label{tab:ragas_results_detailed}
    \resizebox{\textwidth}{!}{%
    \begin{tabular}{|p{0.5\textwidth}|l|l|}
        \hline
        \textbf{Question (Category)} & \textbf{Faithfulness} & \textbf{Answer Relevancy} \\
        \hline
        \textbf{Event-oriented Query:} What were the primary reasons for the start of the American Civil War? & \textbf{0.833} & \textbf{0.826} \\
        \hline
        \multicolumn{3}{|p{0.95\textwidth}|}{
            \textbf{Generated Answer:} Les principales raisons pour le déclenchement de la Guerre civile américaine sont identifiées comme étant la question de l'esclavage et les différences économiques entre le Nord industriel et le Sud agricole. Cependant, les causes de la sécession sont complexes et ont été controversées depuis le début de la guerre. Cet événement a commencé le 12 avril 1861, lorsque l'artillerie de la Caroline du Sud a ouvert le feu sur Fort Sumter.
        } \\
        \hline
        \textbf{Entity-focused Query:} Qui est Antoine Meillet? (French) & \textbf{0.500} & \textbf{1.000} \\
        \hline
        \multicolumn{3}{|p{0.95\textwidth}|}{
            \textbf{Generated Answer:} Antoine Meillet est un linguiste français, comme l'a souligné le linguiste français Antoine Meillet dans l'article "Arménien" (Source ID: 273).
        } \\
        \hline
        \textbf{Event-oriented Query:} Quelles ont été les conséquences majeures de la Première Guerre mondiale en Europe? (French) & \textbf{0.000} & \textbf{0.000} \\
        \hline
        \multicolumn{3}{|p{0.95\textwidth}|}{
            \textbf{Generated Answer:} Je ne peux pas répondre à cette question en me basant uniquement sur les informations fournies. Les extraits de journaux ne fournissent pas de détails sur les conséquences majeures de la Première Guerre mondiale en Europe.
        }\\
        \hline
        \textbf{Unanswerable Query:} Expliquez en détail comment les voyages interstellaires des Romains ont influencé l'architecture des temples égyptiens. & \textbf{0.000} & \textbf{0.000} \\        
        \hline
        \multicolumn{3}{|p{0.95\textwidth}|}{
            \textbf{Generated Answer:} Je ne peux pas répondre à cette question en me basant uniquement sur les informations fournies. Les extraits de journaux fournis ne mentionnent aucun lien entre les voyages interstellaires des Romains et l'architecture des temples égyptiens.
        } \\
        \hline
    \end{tabular}
    }%
\end{table*}

The Time column reports total re-ranking time per model over the 50 queries under both conditions (model load excluded): 5.0\,s for the selected retriever, rising to 50.4\,s for \texttt{Linq-Embed-Mistral}; fusion issues one index search per reformulation. Per language (25 FR, 25 EN), the selected retriever's dense Sim@5 is 0.870 (FR) and 0.859 (EN), and fusion lowers it in both (0.864 and 0.837); per-language means for all four models are released in \texttt{r2\_results.csv}. The overhead is modest, with no compensating similarity gain. Comparing fusion only against single-query dense isolates query expansion (the deployed pipeline fuses multi-query \emph{dense} lists; the appendix additionally evaluates fusing a \emph{sparse} list); under the same protocol, BM25 trails single-query dense by \(0.008\) to \(0.011\) Sim@1 and fusing it into the dense list yields no gain (the hybrid matches dense to within \(0.002\), no interval excluding zero); expansion lowers re-scored similarity for dense, sparse and fused retrieval alike, so the negative result above is a property of the protocol, not of dense retrieval. Full six-condition results and intervals are in \cref{tab:app_sparse_conditions}; alternative fusion operators (\eg, CombSUM) and gold-label recall remain future work.

\subsubsection{QA Quality Assessment and Boundary Conditions}
Under RAGAS (\cref{tab:ragas_results_detailed}), faithfulness averages 0.354 (sd 0.343) and answer relevancy 0.370 (sd 0.430) over the 50 queries; by category, entity-focused queries reach 0.428/0.545, event-oriented 0.406/0.381, and the intentionally unanswerable 0.100/0.000. Abstentions receive zero relevancy by construction and pull the aggregates down: the system refuses 28 of the 50 questions---all 10 unanswerable ones and 18 of the 40 answerable ones. On well-supported queries it scores high: ``{\ttfamily Qui était Victor Hugo ?}'' and ``{\ttfamily Who was Alan Turing?}'' reach perfect faithfulness (1.000) with relevancy 1.000 and 0.862, and the American Civil War query scores 0.833/0.826 (\cref{tab:ragas_results_detailed}). 

Two boundary conditions are visible in \cref{tab:ragas_results_detailed}. By language, faithfulness is comparable (0.356 FR vs.\ 0.351 EN) while relevancy is higher on French queries (0.404 vs.\ 0.337), 25 queries each. Broad synthesis queries trigger abstention rather than speculation: the French WWI-consequences query is refused (0.000/0.000) because its retrieved passages do not carry the cross-event synthesis it demands, and the intentionally unanswerable ``interstellar travels of the Romans'' query yields 0.000/0.000 with an explicit refusal---the fail-safe against misinformation the design targets. The refusal rate on answerable queries (18 of 40) shows the same trade-off at scale: strict grounding keeps every claim traceable to a cited extract, but top-\(K\) sentence-level evidence is often too narrow for broad questions, and coverage rather than fluency is the limiting factor. The pipeline excels on narrow, well-supported queries in both languages, while broad interpretive questions require abstention; confidence scoring and evidence highlighting remain recommended interface enhancements.

\subsection{Discussion}
\label{subsec:discussion}

Our objective for MTRACE was to mitigate vocabulary mismatch for multilingual QA over temporally layered text; the results let us state what this testbed supports. \textbf{First, single-query dense retrieval is the robust default under this protocol.} Query expansion did not improve re-scored similarity for any retriever: dense fusion trails dense retrieval (\cref{tab:benchmark_retrieval_fusion_detail}), and the same holds for sparse and sparse+dense retrieval (\cref{sec:app_sparse}), so the mechanism we set out to exploit did not pay off on this corpus under original-query re-scoring---\cref{subsec:exp_e2e} gives the mechanism (RRF retains reformulation-favoured candidates that the original-query metric then penalises). \textbf{Second, the choice of metric is part of the finding.} What fusion is designed to improve---evidence coverage against gold answers---cannot be measured with 50 qualitative queries, which we flag as the decisive next experiment (\cref{sec:limitations}). \textbf{Third, the pipeline's value lies in generation and abstention rather than retrieval:} structured grounding yields perfect faithfulness on well-supported queries and refuses all ten unanswerable ones (\cref{tab:ragas_results_detailed}). \textbf{Fourth, the scope of the objective is narrower than the title suggests:} temporal drift motivates the testbed but is not measured, because MIRACL carries no reliable date metadata; the claims are about lexical and cross-lingual variation on a comparatively clean, text-only corpus (\cref{sec:limitations}).

\section{Conclusion and Future Work}
\label{sec:conclusion}

We presented MTRACE, a multilingual RAG pipeline integrating SQE with multi-query fusion and characterized its behavior on French and English MIRACL: under our similarity-based re-scoring protocol, fusion trails single-query dense retrieval across four embedding models; structured generation produced grounded answers and appropriate abstention on well-supported and unanswerable queries; and ablation studies showed the importance of syntactic coherence in NER for entity-based enrichment. Scoping our evaluation to a clean, text-only baseline separated these effects from OCR confounds; direct measurement of diachronic drift, broader language coverage, and validation against human judgments remain open problems.

Future work includes extending this text-only foundation to historical archives with layout-aware parsing and OCR-error correction; alternative fusion operators beyond RRF and component-level attribution of the retrieval stack; and evaluation on larger, more diverse multilingual corpora for cross-lingual generalization and retrieval robustness.

\section{Limitations}
\label{sec:limitations}
Our pipeline is evaluated under conditions that delimit its claims. First, the evaluation covers only the French and English subsets of MIRACL, a comparatively clean, text-only corpus: we do not quantify resilience to OCR noise or layout artifacts of raw historical archives, nor do we evaluate the other 16 MIRACL languages. Second, the study does not measure diachronic lexical drift directly---MIRACL passages carry no reliable date metadata, so temporal variation enters only through query and passage content, and drift claims remain hypotheses about motivation rather than measured outcomes. Third, our 50 evaluation queries were authored by the authors rather than drawn from a query log, and we report no significance tests; the query set and per-query results are released to support follow-up analysis. Fourth, automated metrics, particularly RAGAS, are proxies that may not align with human judgments of historical nuance. Finally, query expansion relies on a general-purpose LLM (Mistral-7B), which may lack knowledge of obscure entities, and adds an expansion call per query (five reformulations), with corresponding compute and latency costs.

% Bibliography
% \bibliographystyle{acl_natbib}
\bibliography{egbib} % Make sure your .bib file is named custom.bib or change this line to match your file name

\clearpage
\appendix
This appendix collects material that supports the main text without carrying new claims: the prompt templates (\ref{sec:appendix_prompts}), the reproducibility record (\ref{sec:app_impl}), formal metric definitions (\ref{sec:app_metrics}), the detailed ablations (\ref{sec:app_ner}, \ref{sec:app_emb}), the retrieval-baseline experiments (\ref{sec:app_sparse}), and an index of the released artifacts (\ref{sec:app_artifacts}).

\section{Prompt Templates}
\label{sec:appendix_prompts}
The three prompt templates used by the deployed pipeline are reproduced below.
They are referenced from \cref{subsec:retrieval,subsec:generation}.

\begin{table}[h]
    \centering
    \footnotesize
    \caption{Prompt template for query variation generation.}
    \label{tab:prompt_query_variation}
    {\ttfamily % start monospace
    \begin{tabular}{p{0.93\linewidth}}
        \hline
        \textbf{Query Variation Generation} \\
        \hline
        \textit{Task:} Reformulate the following question in \textit{\{num\_variations\}} different ways for the purpose of efficient search.\\
        \textit{Constraints:}
        \begin{itemize}[leftmargin=*,noitemsep,topsep=2pt]
            \item Preserve the original meaning
            \item Be concise
            \item One reformulation per line
            \item Do not number the reformulations
        \end{itemize}
        \textit{Input:} \textit{\{original\_query\}}\\
        \textit{Output:} Reformulations:\\
        \hline
    \end{tabular}
    } % end monospace
\end{table}

\begin{table}[h]
    \centering
    \footnotesize
    \caption{Structured prompt for answer generation.}
    \label{tab:prompt_qa}
    {\ttfamily % start monospace
    \begin{tabular}{p{0.93\linewidth}}
        \hline
        \textbf{Answer Generation Prompt} \\
        \hline
        \textit{Task:} Answer the question using \textbf{exclusively} the provided "Segment Extracts".\\
        \textit{Constraints:}
        \begin{itemize}[leftmargin=*,noitemsep,topsep=2pt]
            \item Do not use outside knowledge or make assumptions.
            \item If information is missing, state: "I cannot answer this question based solely on the provided information."
            \item Verify that extracted details relate to the main event, not unrelated mentioned events.
            \item Ensure relationships between entities are explicitly described before asserting them.
            \item Do not refer to yourself as an AI model.
            \item Note: A consequence is a result occurring \textit{after} an event; a cause is not a consequence.
        \end{itemize}
        \textit{Input:} : \textit{\{context\_text\}}, Question: \textit{\{query\}}\\
        \textit{Output:} Answer: \{\}\\
        \hline
    \end{tabular}
    } % end monospace
\end{table}

\begin{table}[h]
    \centering
    \footnotesize
    \caption{Structured prompt for text QA.}
    \label{tab:prompt_historical_qa}
    {\ttfamily % start monospace
    \begin{tabular}{p{0.93\linewidth}}
        \hline
        \textbf{Text QA Prompt} \\
        \hline
        You must answer the following question using the provided text excerpts. Your task is to answer the question using EXCLUSIVELY the information contained in the excerpts provided below.\\
        \\
        \textit{Input:} Segment Excerpts: \{\textit{context\_text}\} \\
        \textit{Input:} Question: \{\textit{query}\} \\
        \\
        \textit{Constraints:}
        \begin{itemize}[leftmargin=*,noitemsep,topsep=2pt]
            \item Make no assumptions; do not use any external knowledge.
            \item If the excerpts don't contain the necessary information to answer the question, you MUST explicitly state: "I cannot answer this question based solely on the provided information."
            \item Answer in the same language as the question.
            \item Carefully verify that each piece of information extracted pertains solely to the main event of the question, excluding those mentioned in the context, unless the causal link is explicit.
            \item If you identify actors, ensure their relationships are explicitly described in the excerpts before asserting them.
            \item Do not refer to yourself as an "AI model".
            \item A consequence is a result or effect that occurs after an event. An event that triggers or causes another event is not a consequence of that event itself.
        \end{itemize}
        \textit{Output:} \{\}\\
        \hline
    \end{tabular}
    }
\end{table}

\section{Reproducibility and Implementation Details}
\label{sec:app_impl}

This appendix records the configuration and hardware behind every reported number.

\paragraph{Indexing, retrieval and generation.} The corpus is split into 688{,}992 chunks of 512 tokens and indexed with the dense retriever only (no sparse index is deployed). Retrieval feeds the generator \(K=10\) candidates fused by RRF (\(k=60\)); query expansion issues one generator call per query (\cref{tab:prompt_query_variation}) and answer generation a second (\cref{tab:prompt_qa,tab:prompt_historical_qa}), at temperature 0.3.

\paragraph{Hardware and precision.} Encoding times in \cref{tab:embedding-comprehensive} were measured in the original cluster environment of the model-selection study. Re-ranking times in \cref{tab:benchmark_retrieval_fusion_detail} and answer generation behind \cref{tab:ragas_results_detailed} used a single NVIDIA T4 (16\,GB) in fp16. The retrieval baselines in \cref{sec:app_sparse} were produced in one session on the same T4 class, with the selected retriever scored in fp32 and the three 7B models in 4-bit quantization; that appendix quantifies the precision shift and computes every contrast within a single session.

\paragraph{Judge and embedding stand-ins.} RAGAS claim verification uses \texttt{deepseek-v4-flash} and relevancy embeddings use \texttt{qwen3-embedding:4b}, hosted stand-ins for the framework defaults \texttt{gpt-4o-mini} and \texttt{text-embedding-ada-002} (\cref{sec:app_metrics}); those scores are proxies for human judgment of historical nuance (\cref{sec:limitations}).

\section{Evaluation Metrics: Formal Definitions}
\label{sec:app_metrics}

This appendix gives the formal definitions and protocol details behind \cref{subsec:metrics}.

\paragraph{Retrieval.} Top-\(K\) similarity is the mean cosine similarity between a query embedding and its \(K\) top-ranked re-scored candidates,
\(\text{Sim}@K = \frac{1}{|Q|}\sum_{q \in Q} \frac{1}{K}\sum_{i=1}^{K} \cos\!\left(q, d_i^{(q)}\right)\),
with \(Q\) the query set and \(d_i^{(q)}\) the \(i\)-th ranked candidate for \(q\). In the self-retrieval variant used for embedding selection, every corpus chunk queries the full-corpus index including itself, so the rank-1 hit is the chunk itself (cosine \(1\)) and the reported Top-5 quantity is
\(\text{Top5} = \frac{1}{|\mathcal{C}|}\sum_{d \in \mathcal{C}} \frac{1}{5}\left(1 + \sum_{i=1}^{4} \cos(d, d_i^{(d)})\right)\);
the confidence drop is the top-2 gap \(\Delta_{\text{conf}} = s(d_1) - s(d_2)\), with \(s(d_i)\) the score of the \(i\)-th ranked text. The constant self-match shifts both self-retrieval quantities identically for every model, so cross-model comparisons in \cref{tab:embedding-comprehensive} remain valid, though the absolute values are inflated relative to a leave-one-out protocol; query-based scores in \cref{tab:benchmark_retrieval_fusion_detail} are not directly comparable to self-retrieval scores.

\paragraph{NER.} Syntactic relevance is the proportion of syntactically coherent extracted entities,
\(\text{SynRel} = \frac{1}{\sum_i |E_i|}\sum_i \sum_{e \in E_i} \mathbb{1}_{\text{coherent}}(e)\),
where the indicator marks entities that form a complete linguistic unit without fragmentation or boundary errors.

\paragraph{Generation (RAGAS).} Faithfulness is the fraction of atomic claims in an answer \(a\) supported by the retrieved context \(R\),
\(\text{Faithfulness}(a, R) = |\text{claims}(a)|^{-1} \sum_{c \in \text{claims}(a)} \mathbb{1}_{\text{supported}}(c, R)\).
Answer relevancy generates \(N\) synthetic questions \(g_i\) from \(a\) with an LLM and averages their cosine similarity to the original query,
\(\text{Relevancy}(a,q) = \frac{1}{N}\sum_{i=1}^{N} \cos(E_{g_i}, E_o)\), with \(E_x = \text{embed}(x)\).
In the reported runs, claim verification and synthetic-question generation use \texttt{deepseek-v4-flash} and relevancy embeddings use \texttt{qwen3-embedding:4b}, hosted stand-ins for the framework defaults \texttt{gpt-4o-mini} and \texttt{text-embedding-ada-002}~\citep{es2024ragas}; we report scores as proxies and note in \cref{sec:limitations} that they may not align with human judgments of historical nuance.

\paragraph{Clustering diagnostics.} Silhouette, Davies--Bouldin and Calinski--Harabasz are standard vector-space diagnostics (higher Silhouette and Calinski--Harabasz, lower Davies--Bouldin, indicate better-separated clusters). \cref{tab:embed-cluster} reports them for the five candidate models; \cref{subsec:exp_embedding} states how we interpret them---as a weak negative filter, not as evidence of well-separated semantic structure.

\section{NER Ablation: Detailed Results}
\label{sec:app_ner}

\begin{figure*}[t]
    \centering
    \footnotesize
    \begin{subfigure}[b]{0.32\linewidth}
        \centering
        \includegraphics[width=\linewidth]{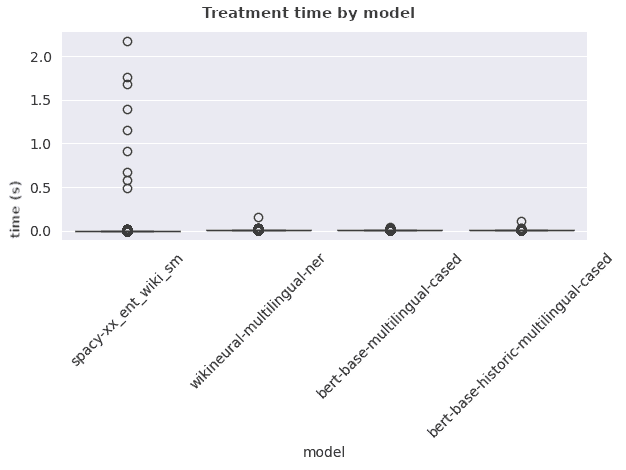}
        \caption{Processing time}
        \label{fig:ner-speed}
    \end{subfigure}
    \hfill
    \begin{subfigure}[b]{0.32\linewidth}
        \centering
        \includegraphics[width=\linewidth]{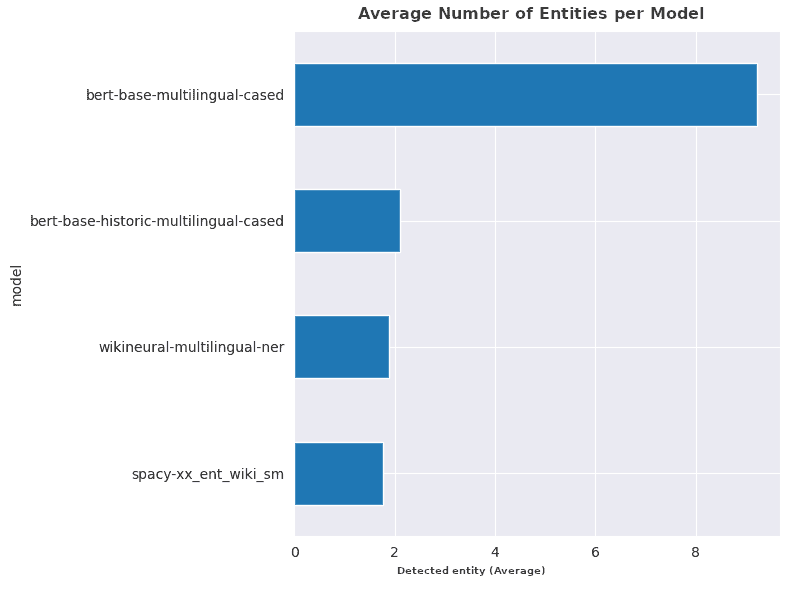}
        \caption{Entity detection rate}
        \label{fig:ner-entities}
    \end{subfigure}
    \hfill
    \begin{subfigure}[b]{0.32\linewidth}
        \centering
        \includegraphics[width=\linewidth]{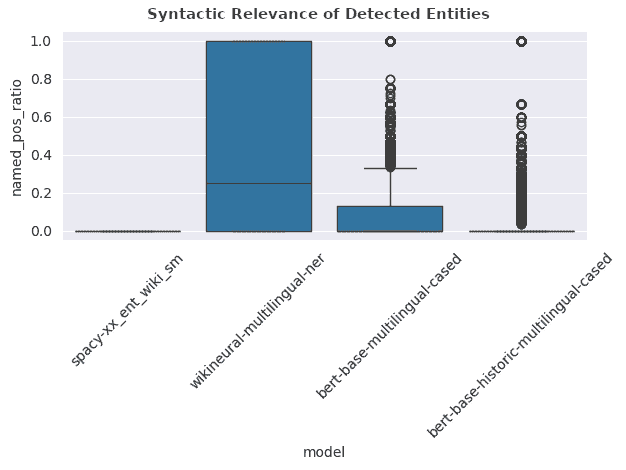}
        \caption{Syntactic relevance}
        \label{fig:ner-syntax}
    \end{subfigure}
    \caption{Comparative evaluation of NER models on multilingual text, assessing: (a) computational efficiency; (b) extraction capability; and (c) output quality (syntactic coherence).}
    \label{fig:ner-comparison}
\end{figure*}

\cref{tab:ner-qualitative-comparison} gives the qualitative comparison behind the NER selection; \cref{fig:ner-comparison} shows the same three criteria visually---processing time, entity detection rate and syntactic relevance. The ranking agrees with the qualitative table, and \cref{subsec:exp_ner} draws the selection conclusion from it.

\section{Embedding Model Ablation: Detailed Results}
\label{sec:app_emb}

\begin{figure*}[t]
    \centering
    \begin{subfigure}[b]{0.32\linewidth}
        \centering
        \includegraphics[width=\linewidth]{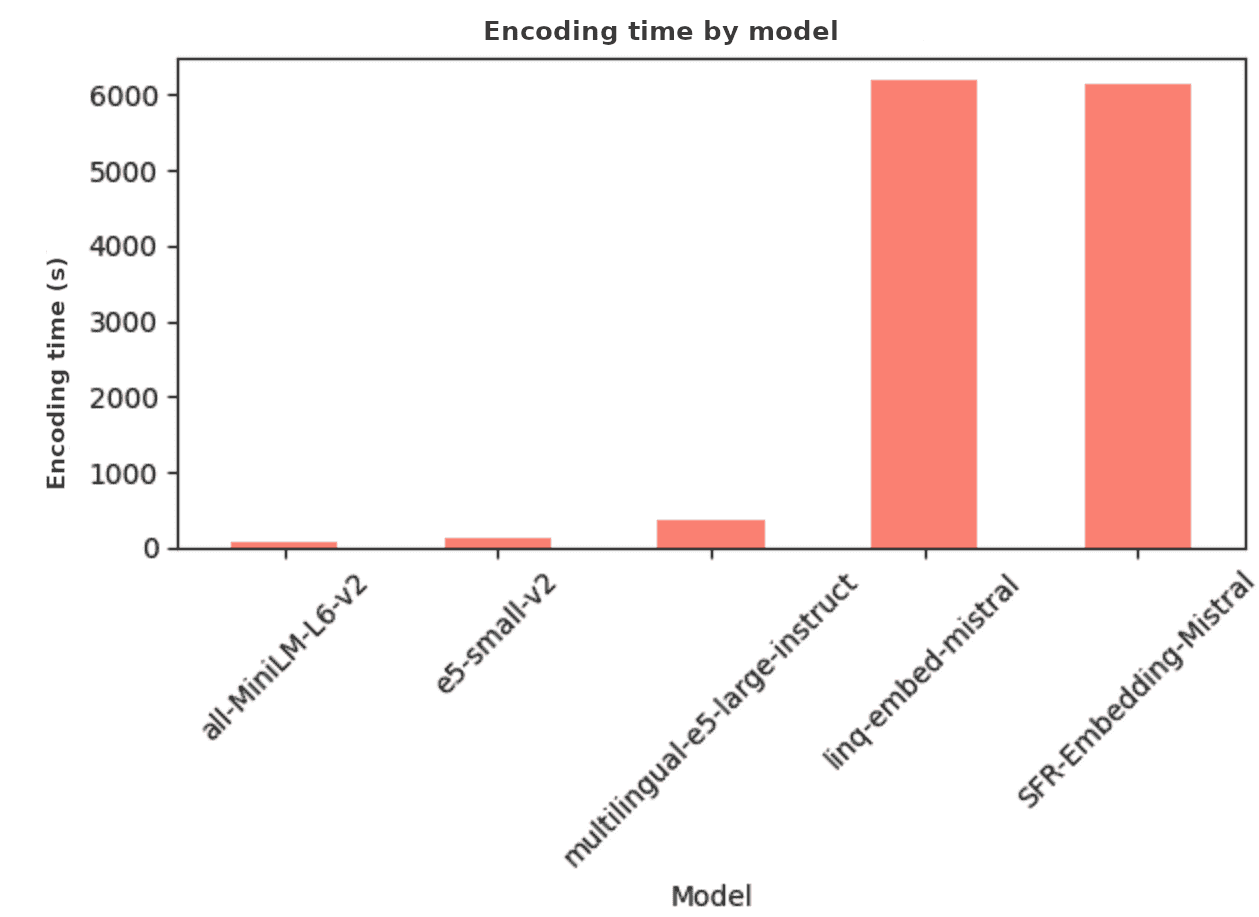}
        \caption{Encoding time}
        \label{fig:embed-time}
    \end{subfigure}
    \hfill
    \begin{subfigure}[b]{0.32\linewidth}
        \centering
        \includegraphics[width=\linewidth]{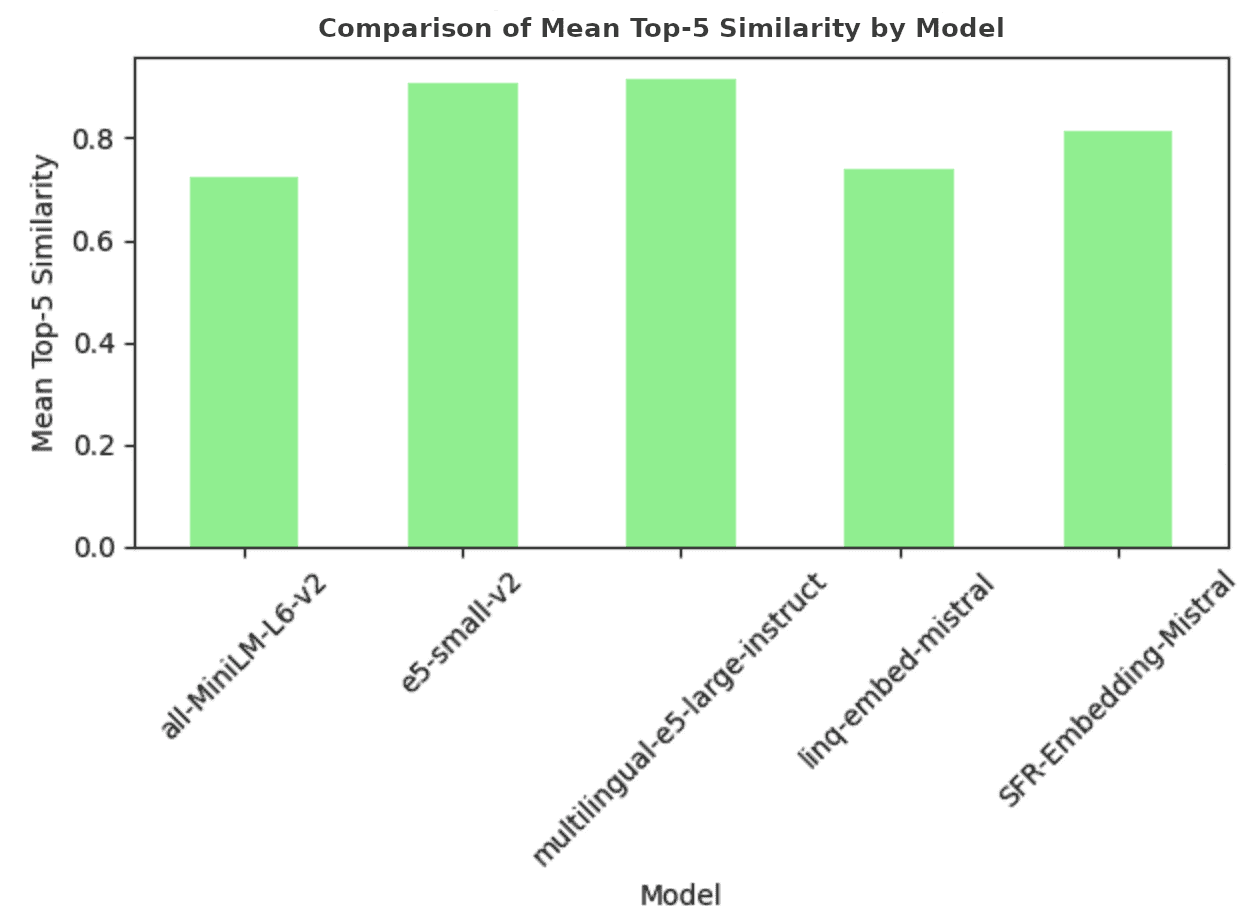}
        \caption{Top-5 similarity rate}
        \label{fig:embed-top5}
    \end{subfigure}
    \hfill
    \begin{subfigure}[b]{0.32\linewidth}
        \centering
        \includegraphics[width=\linewidth]{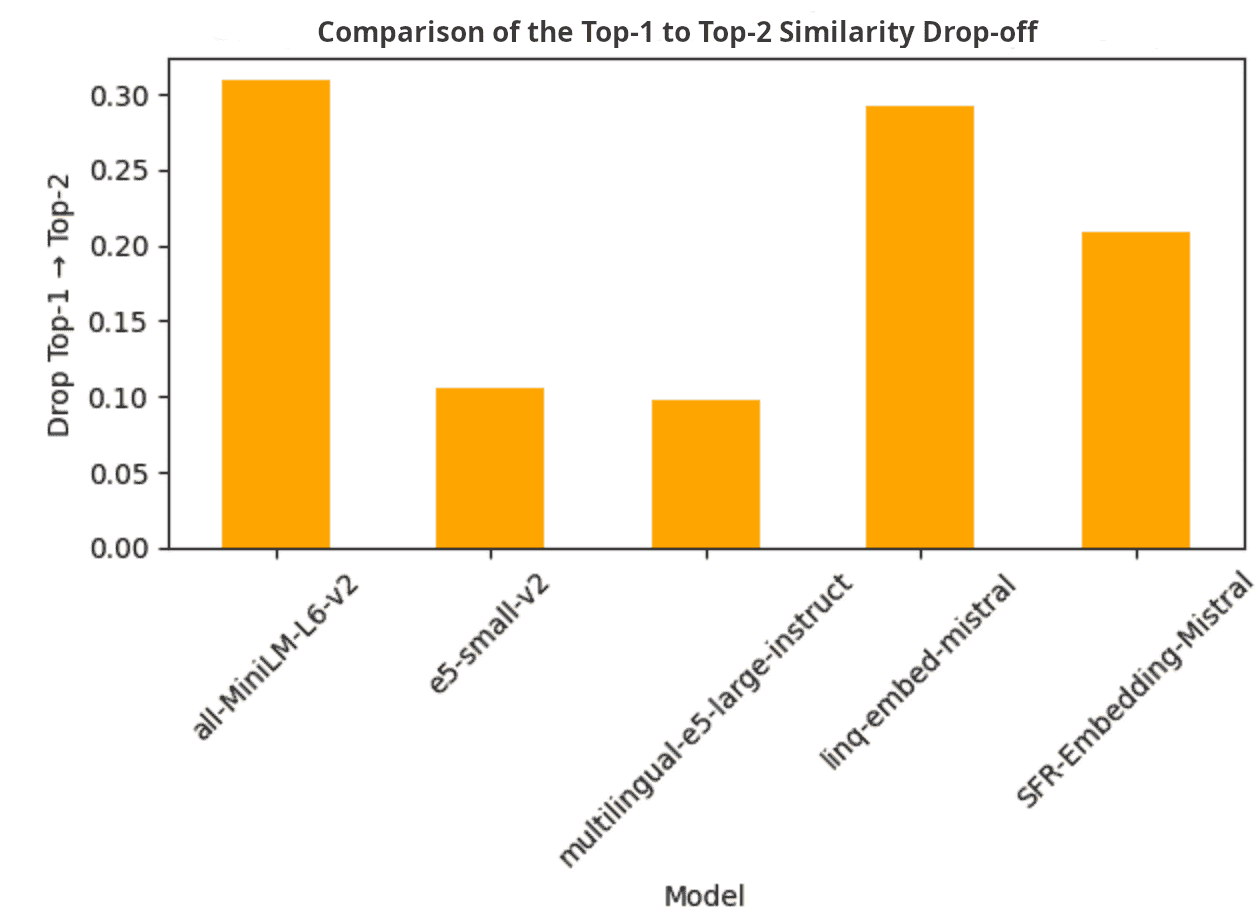}
        \caption{Confidence drop}
        \label{fig:embed-drop}
    \end{subfigure}
    \caption{Evaluation of embedding models: (a) computational efficiency; (b) semantic retrieval performance; and (c) retrieval confidence.}
    \label{fig:embedding-evaluation}
\end{figure*}

\cref{tab:embedding-comprehensive} reports the numbers behind the retriever selection; \cref{fig:embedding-evaluation} shows the same three dimensions visually: computational efficiency, semantic retrieval performance, and retrieval confidence. The ranking in the table and the panels agree, and \cref{subsec:exp_embedding} draws the selection conclusion from them.

\section{Sparse and Hybrid Retrieval Baselines}
\label{sec:app_sparse}

\begin{table*}[t]
    \centering
    \footnotesize
    \caption{Six retrieval conditions under the protocol of \cref{tab:benchmark_retrieval_fusion_detail}: mean cosine Sim@1\,/\,Sim@5 after re-scoring each condition's top-10 candidates with the listed model, over the 50 queries. Dense and Dense+exp replicate the two conditions of \cref{tab:benchmark_retrieval_fusion_detail} within one session and one scorer precision (see provenance below). Hybrid fuses the dense and BM25 lists with RRF (\(k=60\)). Model abbreviations: \texttt{e5-large} = \texttt{multilingual-e5-large}, \texttt{e5-mistral} = \texttt{e5-mistral-7b-instruct}, \texttt{SFR} = \texttt{SFR-Embedding-Mistral}, \texttt{Linq} = \texttt{Linq-Embed-Mistral}.}
    \label{tab:app_sparse_conditions}
    \begin{tabular}{|l|cccc|}
        \hline
        Condition & \texttt{e5-large} & \texttt{e5-mistral} & \texttt{SFR} & \texttt{Linq} \\
        \hline
        Dense        & 0.8755\,/\,0.8647 & 0.6880\,/\,0.5780 & 0.7277\,/\,0.6941 & 0.6171\,/\,0.5785 \\
        Dense+exp    & 0.8657\,/\,0.8511 & 0.5949\,/\,0.4701 & 0.7056\,/\,0.6689 & 0.5882\,/\,0.5459 \\
        \hline
        BM25         & 0.8679\,/\,0.8482 & 0.6802\,/\,0.5850 & 0.7168\,/\,0.6693 & 0.6059\,/\,0.5579 \\
        BM25+exp     & 0.8579\,/\,0.8400 & 0.6294\,/\,0.5250 & 0.6972\,/\,0.6538 & 0.5810\,/\,0.5378 \\
        \hline
        Hybrid       & 0.8754\,/\,0.8635 & 0.6879\,/\,0.5870 & 0.7279\,/\,0.6911 & 0.6190\,/\,0.5777 \\
        Hybrid+exp   & 0.8676\,/\,0.8534 & 0.6303\,/\,0.5137 & 0.7117\,/\,0.6757 & 0.5982\,/\,0.5586 \\
        \hline
    \end{tabular}
\end{table*}

\begin{table*}[t]
    \centering
    \footnotesize
    \caption{Paired per-query differences in cosine Sim@1 under the single-session protocol above: each cell gives the mean difference and, beneath it, the bootstrap 95\% CI (10\,000 resamples, seed 42). \textsuperscript{*} marks intervals excluding zero. Dense+exp $-$ Dense reproduces the effect in \cref{tab:benchmark_retrieval_fusion_detail}; the final two rows give Sim@5.}
    \label{tab:app_deltas}
    \begin{tabular}{|l|l|l|l|l|}
        \hline
        \textbf{Contrast} & \texttt{e5-large} & \texttt{e5-mistral} & \texttt{SFR} & \texttt{Linq} \\
        \hline
        BM25 $-$ Dense & \shortstack{$-0.0076^{*}$\\$[-0.0112,-0.0045]$} & \shortstack{$-0.0078$\\$[-0.0298,+0.0141]$} & \shortstack{$-0.0109^{*}$\\$[-0.0178,-0.0045]$} & \shortstack{$-0.0112^{*}$\\$[-0.0198,-0.0030]$} \\
        Hybrid $-$ Dense & \shortstack{$-0.0001$\\$[-0.0002,+0.0000]$} & \shortstack{$-0.0001$\\$[-0.0125,+0.0124]$} & \shortstack{$+0.0001$\\$[-0.0018,+0.0022]$} & \shortstack{$+0.0019$\\$[-0.0018,+0.0058]$} \\
        Hybrid $-$ BM25 & \shortstack{$+0.0075^{*}$\\$[+0.0044,+0.0111]$} & \shortstack{$+0.0078$\\$[-0.0094,+0.0250]$} & \shortstack{$+0.0110^{*}$\\$[+0.0054,+0.0172]$} & \shortstack{$+0.0131^{*}$\\$[+0.0063,+0.0206]$} \\
        Dense+exp $-$ Dense & \shortstack{$-0.0098^{*}$\\$[-0.0143,-0.0058]$} & \shortstack{$-0.0931^{*}$\\$[-0.1289,-0.0599]$} & \shortstack{$-0.0221^{*}$\\$[-0.0325,-0.0130]$} & \shortstack{$-0.0289^{*}$\\$[-0.0423,-0.0166]$} \\
        BM25+exp $-$ BM25 & \shortstack{$-0.0100^{*}$\\$[-0.0167,-0.0040]$} & \shortstack{$-0.0508^{*}$\\$[-0.0893,-0.0162]$} & \shortstack{$-0.0196^{*}$\\$[-0.0345,-0.0061]$} & \shortstack{$-0.0250^{*}$\\$[-0.0417,-0.0097]$} \\
        Hybrid+exp $-$ Hybrid & \shortstack{$-0.0078^{*}$\\$[-0.0122,-0.0039]$} & \shortstack{$-0.0577^{*}$\\$[-0.0917,-0.0260]$} & \shortstack{$-0.0161^{*}$\\$[-0.0251,-0.0083]$} & \shortstack{$-0.0208^{*}$\\$[-0.0312,-0.0114]$} \\
        Sim@5: BM25 $-$ Dense & \shortstack{$-0.0165^{*}$\\$[-0.0201,-0.0131]$} & \shortstack{$+0.0070$\\$[-0.0091,+0.0234]$} & \shortstack{$-0.0248^{*}$\\$[-0.0325,-0.0177]$} & \shortstack{$-0.0206^{*}$\\$[-0.0281,-0.0135]$} \\
        Sim@5: Hybrid $-$ Dense & \shortstack{$-0.0012^{*}$\\$[-0.0017,-0.0007]$} & \shortstack{$+0.0089$\\$[-0.0016,+0.0195]$} & \shortstack{$-0.0029^{*}$\\$[-0.0059,-0.0003]$} & \shortstack{$-0.0008$\\$[-0.0038,+0.0021]$} \\
        \hline
    \end{tabular}
\end{table*}

This appendix documents a controlled comparison against the two retrieval baselines deferred from
\cref{subsec:exp_e2e}: a sparse (BM25) retriever and a sparse+dense fusion variant. Both are
\emph{evaluated alternatives}, not the deployed configuration---the deployed index is dense only
(\cref{subsec:data_preprocessing})---and they are reported here so that the contribution of a
sparse leg is measured rather than assumed.

\subsection{Scope and protocol}
Every condition follows the protocol of \cref{tab:benchmark_retrieval_fusion_detail}: the same
corpus, the same 50 evaluation queries, the same cached reformulations, top-\(K=10\) candidates and
RRF with \(k=60\), re-scored by the same four models against the original query. Dense retrieval is
exact cosine over the precomputed embedding matrix---no approximate index is involved---and BM25
uses Lucene defaults (\(k_1=1.5\), \(b=0.75\)). \cref{tab:app_sparse_conditions} lists the six
resulting conditions and their Sim@1/Sim@5 means.

\subsection{Results}
\cref{tab:app_deltas} reports the paired per-query contrasts behind these means. Three patterns are
visible. First, dense retrieval exceeds the sparse baseline, but modestly: BM25 trails dense by
\(0.008\) to \(0.011\) Sim@1---with confidence intervals excluding zero for three of the four
models---and by \(0.017\) to \(0.025\) Sim@5. Second, fusing a sparse leg into the dense list
does not help: Hybrid matches Dense to within \(0.002\) Sim@1 for every model, no interval
excluding zero, and stays within \(0.003\) Sim@5; the fused list is therefore no better than
dense retrieval alone and clearly better than sparse retrieval alone, so the sparse leg contributes
nothing beyond the dense list it is merged with. Third, query expansion is the one factor that moves
all three retrievers in the same direction: it lowers re-scored similarity for dense
(Dense+exp $-$ Dense), for sparse (BM25+exp $-$ BM25) and for their fusion
(Hybrid+exp $-$ Hybrid), with every interval excluding zero. This extends the main text's
dense-only observation to two further retrieval paradigms and supports the same mechanism: RRF
retains candidates that a reformulation favours, while the metric re-scores against the original
query.

\subsection{Precision control}
All six conditions come from one session on a single NVIDIA T4, with the selected retriever scored
in fp32 and the three 7B models in 4-bit quantization, whereas
\cref{tab:benchmark_retrieval_fusion_detail} was measured in fp16. We therefore re-ran that table's
two conditions in the same session and precision (the Dense and Dense+exp rows of
\cref{tab:app_sparse_conditions}) and compared them query by query. The shift is \(+0.0001\)
Sim@1 for the selected retriever but \(+0.007\) to \(+0.009\) Sim@1 for the 4-bit models,
reaching \(+0.012\) in Sim@5---larger than most effects in \cref{tab:app_deltas}---so every
contrast we report is computed within a single session with one scorer precision, and no score is
compared across precisions. The same control reproduces the main text's fusion effect closely
(\(-0.0098\), \(-0.0931\), \(-0.0221\), \(-0.0289\) versus \(-0.010\), \(-0.087\),
\(-0.022\), \(-0.030\), in the order of \cref{tab:benchmark_retrieval_fusion_detail}), which
independently validates the numbers reported in \cref{subsec:exp_e2e}.

\subsection{Caveats}
(i) Dense and Hybrid candidate lists partly originate from the encoder that later re-scores them,
while BM25 lists do not, so the protocol structurally favours dense-origin candidates---the same
asymmetry discussed in \cref{subsec:exp_e2e}. (ii) Fifty queries from a single seed: we report no
variance across seeds or reformulation draws, so effects smaller than roughly \(0.01\) should be
read as indicative rather than established. (iii) In 4-bit, Sim@1 for \texttt{e5-mistral-7b-instruct}
has visibly wider intervals and two of its contrasts remain inconclusive. The per-query results and
summaries are released with the artifact bundle, together with the run script.

\section{Released Artifacts}
\label{sec:app_artifacts}

The anonymous repository linked in the abstract contains the following, so every reported number can be traced or regenerated:

\begin{itemize}
    \item \texttt{queries\_50.csv}, \texttt{verify\_queries.py}---the 50 evaluation queries and their corpus-grounding check (\cref{subsec:data_preprocessing}).
    \item \texttt{r2\_results.csv} and \texttt{r2\_summary.json}---the dense vs.\ fusion retrieval benchmark per query and model (\cref{tab:benchmark_retrieval_fusion_detail}).
    \item \texttt{r3\_answers.json} and \texttt{ragas\_evaluation\_results.csv}---generated answers and RAGAS scores (\cref{tab:ragas_results_detailed}).
    \item \texttt{dense\_*}, \texttt{bm25\_*} and \texttt{hybrid\_*} result files together with the run script---the dense replication and the sparse and sparse+dense baselines (\cref{sec:app_sparse}).
    \item The three prompt templates---the verbatim prompts used by the deployed pipeline (\cref{tab:prompt_query_variation,tab:prompt_qa,tab:prompt_historical_qa}).
\end{itemize}

\end{document}